\documentstyle[prl,aps]{revtex}
\draft
\begin{document}
\twocolumn[\hsize\textwidth\columnwidth\hsize\csname
@twocolumnfalse\endcsname

\title{Pearling Instabilities of Membrane Tubes with Anchored Polymers}
\author{Ilan Tsafrir$^1$, Dror Sagi$^1$, Tamar Arzi$^1$, Marie-Alice
Guedeau-Boudeville,$^2$ \\
Vidar Frette$^{1,}$\cite{vidar},  Daniel Kandel$^{1}$,
and Joel Stavans$^1$}
\address{$^1$ Department of Physics of Complex Systems, \\
The Weizmann Institute of Science,
Rehovot 76~100, Israel \\
$^2$ Laboratoire de Physique de la Mati\`{e}re Condens\'{e}e, URA 792,
Coll\`{e}ge de France \\
11 Place Marcelin Berthelot,
F-75231 Paris CEDEX 05, France \\}
      \maketitle

\begin{abstract}
We have studied the pearling instability induced on hollow tubular 
lipid vesicles by hydrophilic polymers with hydrophobic side groups 
along the backbone. The results show that the polymer concentration 
is coupled to local membrane curvature. The relaxation of a pearled 
tube is characterized by two different well-separated time scales, 
indicating two physical mechanisms. We present a model, which 
explains the observed phenomena and predicts polymer segregation 
according to local membrane curvature at late stages.
\end{abstract}
\vspace{0.3cm}
{\hspace*{2.cm}}PACS numbers: 87.16.Dg, 68.10.-m
\vspace{0.3cm}
]
\input epsf
\newpage

Single-component phospholipid membranes have been the focus of intense 
interest in recent years, as the simplest model system of biological 
membranes \cite{Seifert:95}.  The latter are highly complex systems, 
comprised of a bilayer consisting of many types of lipids as well as a 
mesh of macromolecules such as proteins and polysaccharides, which 
participate in a large variety of cellular processes.  A natural step 
to mimic this complexity in a simple model system is to study the 
association of polymeric molecules with self-assembled 
single-component phospholipid membranes\cite{Ringsdorf:88}.

Experiments with polymers, which associate with membranes by {\it 
anchoring}, have revealed changes in the bending moduli of bilayers 
of single-tailed surfactants \cite{Yang:98}, and striking 
morphological changes in 
vesicles\cite{Ringsdorf:88,Simon:95,frette:99}. Anchoring occurs by 
the penetration of a number of hydrophobic side-groups grafted along a 
hydrophilic backbone into a bilayer. Hollow tubular vesicles 
incubated in a solution of anchoring polymers having a polysaccharide 
backbone develop a pearling instability, above a threshold polymer 
concentration\cite{Ringsdorf:91}. This instability was not 
observed with purely hydrophilic polymers, and was effected by 
hydrophobic groups alone (without the backbone), but only at 
concentrations five orders of magnitude higher than with anchoring 
polymers. 

In this experimental and theoretical study we investigate the mechanisms
responsible for pearling in our system. 
It has been suggested that the induction of curvature by the 
anchors, which sink into the membrane to a depth of half a bilayer, 
may constitute a mechanism which drives the pearling instability 
\cite{Ringsdorf:88}. We present a novel nonequilibrium experiment, which shows
that two
independent mechanisms -- spontaneous curvature and area difference --
contribute to the pearling phenomenon. We also present new experimental and 
theoretical findings showing inhomogeneous shapes at late stages.

In our experiments, vesicles were made of
stearoyl-oleoyl-phosphatidyl-choline 
(SOPC) with \(C_{18}\) alkyl chains. The polymer used was hydrophilic 
dextran with a molecular weight of $162,000\,g/mol$, functionalized both 
with palmitoyl alkyl chains \(C_{16}\) and dodecanoic NBD chains as fluorescent 
markers. The anchors are distributed statistically along the backbone, spaced
four persistence lengths apart on average
(1 alkyl chain per 25 glucose units). $1\,\mu l$ droplets of SOPC in a 7:1
chloroform-methanol 
solution ($10\,mg/ml$) were placed on a glass coverslip forming small 
lipid patches. The sample was prehydrated for 20 minutes under 
water-saturated nitrogen, then hydrated with $0.1\,mM$ potassium buffer 
at pH 6.5. A large number of hollow tubes with one or more lamellae 
formed after hydration, connected to the lipid patch. The latter 
constitutes a reservoir with which the tubes can exchange lipid 
molecules. Experiments were conducted at room temperature, insuring 
that the membranes were in a fluid-like state.  This allowed free 
diffusion of both lipids and anchored polymers along the bilayers.  
A drop of polymer solution of a given concentration was introduced 
through one of the cell sides. The polymer was added after the tubes 
were formed, and we therefore assume that it anchored mostly on the 
outermost leaflet of the membrane.  Its concentration {\it on} the 
membrane grew from zero as more and more chains anchored from 
solution.  Events were observed by phase contrast and fluorescence 
microscopy and recorded on video.  The NBD markers 
were excited with an argon laser ($488\,nm$),
and observed with a CCD camera.


Fig.\ 1 shows a tube undergoing pearling after 
addition of polymer. The instability typically starts near the end 
cap of the tube (Fig.\ 1a), and gradually propagates along the axis 
(Fig.\ 1b).  This is presumably since the polymer, diffusing from the 
side of the chamber, first reaches the tip of the tube.  Existing 
pearls become gradually more spherical (Fig.\ 1c).  The rate at which 
pearls form depends on the number of lamellae in the walls of the 
tube, as well as on the concentration of polymer in the water ($2 - 
100\,mg/l$), varying from several seconds to many minutes.  A salient 
feature of Fig.\ 1c is the increase in pearl size towards the end of the tube.
This size gradient becomes much more pronounced at very 
long times, as Fig.\ 1d illustrates for another vesicle.  The string 
of pearls separates into a set of small, nearly-uniform 
spheres connected to a group of much larger spheres.

\begin{figure}[h]
	\epsfxsize=70mm
	\centerline{\epsffile{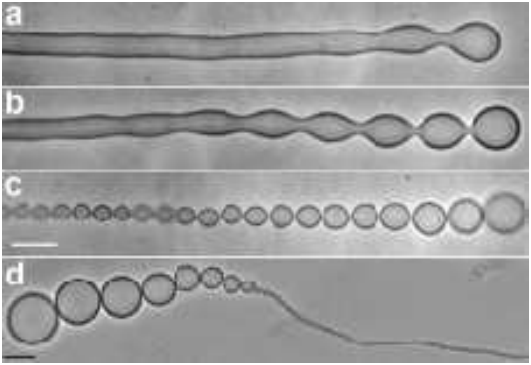}}
	\vspace{0.5cm}
	\caption{Snapshots of a multilamellar tubular vesicle undergoing a pearling
	instability: (a) 0, (b) 70 sec.\ and 
	(c) 150 sec. after onset of pearling. The concentration of
	polymer is below $2\,mg/l$.  (d) Image of an inhomogeneous pearled
	structure at late stages of the instability, 900 sec. after
	onset of pearling. The polymer concentration is below $100\,mg/l$,
	but much larger than in (a-c). The scalebars represent $20\,\mu m$.}
	\label{fig:composite-pearls}
\end{figure}

We have studied the onset of pearling by measuring the wavelength of the 
fastest growing mode, $P$, just above threshold, as 
a function of the radius of the unpearled tube, $R_{0}$.  Within 
experimental accuracy $P$ is linear in $R_0$, with a slope $k\equiv 
2\pi R_0/P=1.07 \pm 0.05$.  This result agrees well both with the 
value $k=1$ predicted by theoretical models based on induced curvature 
as the driving mechanism \cite{Nelson:95}, and with experimental 
results on pearling instabilities in single-tailed surfactant 
systems\cite{Chaieb:98}.  In tension-induced pearling $k\approx 0.7$ 
\cite{Bar-ziv:94}, and thus our measurement excludes pearling due to 
polymer or flow-induced tension in our system.


Polymers anchoring to one side of a membrane can induce 
curvature by two mechanisms. The first is an increase in the area of 
the outermost monolayer into which anchors sink \cite{Mui:95}.  The 
second mechanism is a local deformation of the membrane which can be 
induced either by the anchors regarded as inclusions 
\cite{Leibler:86}, or by an entropic 
pressure exerted by the polymer backbone \cite{Lipowsky:95}. These mechanisms
form the basis of two models 
that describe the tendency of a membrane to display curvature
\cite{Seifert:91}: the 
area difference elasticity model (ADE) and the spontaneous curvature 
model (SC), respectively.  Calculations of equilibrium shapes of
vesicles with 
cylindrical symmetry based on both the SC and ADE models indeed yield 
pearled shapes of constant mean curvature called {\em Delaunay 
surfaces} \cite{Deuling:77}. The equilibrium shape 
is not sensitive to the pearling mechanism.

Nevertheless, the dynamics of pearling may allow us to distinguish 
between the two mechanisms and reveal their presence, since they are
characterized by 
well-separated relaxation time scales.  Inhomogeneities in both area 
difference and spontaneous curvature decay diffusively 
\cite{Seifert:93b}.  The relaxation of spontaneous curvature is 
associated with polymer diffusion in the membrane.  The relevant 
diffusion constant was measured for various macromolecules and falls in 
the range of $1\,\mu m^2/sec<D_{sc}< 5\,\mu m^2/sec$ 
\cite{Almeida:95}.  Area difference relaxes via the sliding of one monolayer
with 
respect to the other, and does not involve diffusion of molecules over 
large distances.  The diffusion constant associated with ADE can be 
estimated from dimensional analysis to be $D_{ade}\approx K_0/b$, 
where $K_0$ is the compression modulus of the membrane, and $b$ is the 
friction coefficient between the two leaflets of a bilayer.  
Experimental estimates of these parameters lead to $50\,\mu 
m^2/sec<D_{ade}< 500\,\mu m^2/sec$ \cite{Seifert:93b}.  Changes in the 
shape of a vesicle are coupled to motion of the surrounding water, 
where the energy dissipated in the water is of the order of the curvature energy
of the bilayer.  From dimensional analysis we find that for both 
mechanisms, ADE and SC, the diffusion constant 
associated with the displacement of water is $D_{hyd} \sim \kappa / 
\eta R_{0} \simeq 50 \mu m^{2}/sec$, where $\kappa$ is the bending modulus 
of the membrane and $\eta$ is the viscosity of 
water. Thus, it is difficult to differentiate between hydrodynamical and ADE
effects by measuring the diffusion constant, but it should be easy to
distinguish between the SC and ADE mechanisms.

To study the relaxational dynamics, a micropipette was used 
to deliver locally a small volume ($\sim 10^{-4}\,\mu l$) of polymer 
solution of concentration $20-100\, mg/l$ close to a tube (Fig.\ 2).
Fluorescence images show that the polymer was indeed 
concentrated in a local region shortly after the injection. Pearling 
occurred, after an induction time in which enhanced undulations were 
observed (Figs.\ 2a-2d). The injection of the polymer was 
stopped after the pearled region increased to a length of order
$100\,\mu m$. The pearled region then shrank gradually (Figs.\ 2e-2h)
as pearls opened up one at a time starting with those farthest from
the region of polymer injection. This decay process took several 
minutes. Fluorescence images show that the amount of polymer in the 
surrounding water is negligible during the decay.
 
We measured the length of the pearled region, $L$, as a function of 
time during the decay process.  For diffusive decay, we expect 
$L^2\approx 2D(\bar{t}-t)$, 
where $D$ is the relevant diffusion constant and $\bar{t}$ is 
a constant.  Figure 3 is a typical example of the dependence of $L^2$ on 
time (measurements on several tubes yielded similar results). The system
exhibits diffusive behavior at early times with a 
diffusion constant of $D\approx 200\,\mu m^2/sec$. We attribute this decay to a
combination of ADE and hydrodynamical effects. At later times 
there is a sharp crossover to a much slower diffusive behavior with 
$D\approx 5\,\mu m^2/sec$. According to our estimates, this 
corresponds to polymer diffusion, and is associated with the SC mechanism. Our
results thus provide clear evidence that both ADE and SC mechanisms influence
the pearling instability.

\begin{figure}[h]
	\epsfxsize=70mm
	\centerline{\epsffile{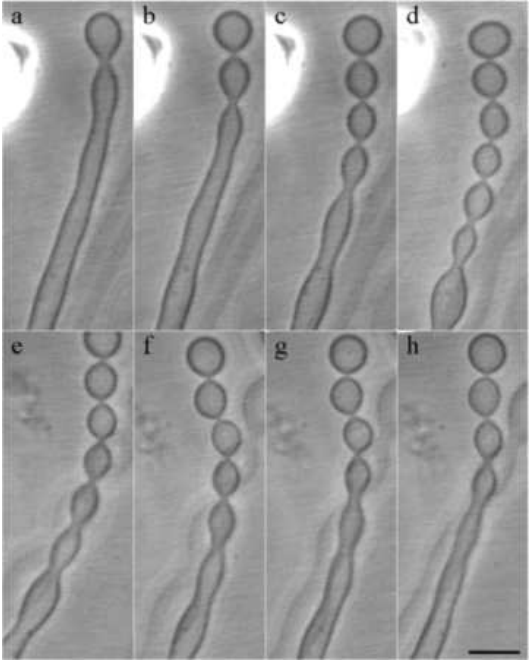}}
	\vspace{0.5cm}
	\caption{Snapshots of a local pearling experiment. (a-d) show the 
	formation of pearls as polymer is added from the micropipette at 
	left. (e-h) show the subsequent opening of these pearls as polymer 
	diffuses down the stem. Times are 2, 5, 8, 11, 42, 46, 54, and 
	65 seconds after the pipette is brought to the vicinity of the 
	vesicle.  The scalebar represents $10\,\mu m$.}
	\label{fig:local-pearls}
\end{figure}


We now carry out a theoretical analysis of the pearling phenomena 
in the case of global application of the polymer 
(Fig.\ 1). We consider closed vesicles with polymer molecules 
only on the outer side of the vesicle, 
and assume that pearling is a result of the SC mechanism (All our 
predictions, except for the inhomogeneities in 
polymer concentration, apply equally well to the ADE model).  In 
contrast with standard curvature models \cite{Seifert:91}, 
the spontaneous curvature in our system is a {\em local} quantity. We assume
that the spontaneous curvature is proportional to 
the polymer concentration on the membrane, $\rho 
H_0$. $\rho(\vec{r})$ is the fraction of the membrane area 
covered by polymer molecules at the position $\vec{r}$, and takes 
values in the interval $[0,1]$.  $H_{0}$ is the spontaneous curvature 
induced by full coverage of the polymer.  We further assume that the 
observed vesicle shapes are close to the equilibrium shapes under the 
constraints of constant vesicle volume and membrane area.  These 
shapes can be obtained by minimizing the free energy of the membrane 
in the presence of the polymer. The simplest free energy for our system 
is a sum of the curvature energy and the entropy of mixing of 
the polymer:

\begin{eqnarray}
F= \int &dA&\left\{2\kappa (H-\rho H_{0})^{2} \right.+ \nonumber\\
&\frac{k_{B}T}{a^{2}}&\left.\left[ \rho \ln\rho + (1-\rho)\ln(1-\rho) 
\right]\right\}~,
\label{eq:free}
\end{eqnarray}
where $H$ is the local mean curvature, $a$ is the characteristic linear size of
an anchored 
polymer molecule and the integration is over the area of the 
membrane. In principle, the effects of gradients of polymer concentration should
also be included in the free energy. However, the inhomogeneities
at equilibrium do not induce an extensive free energy increase (see below), and
we therefore ignore such terms. 

\begin{figure}[h]
	\vspace{0.25cm}
	\epsfxsize=70mm
	\centerline{\epsffile{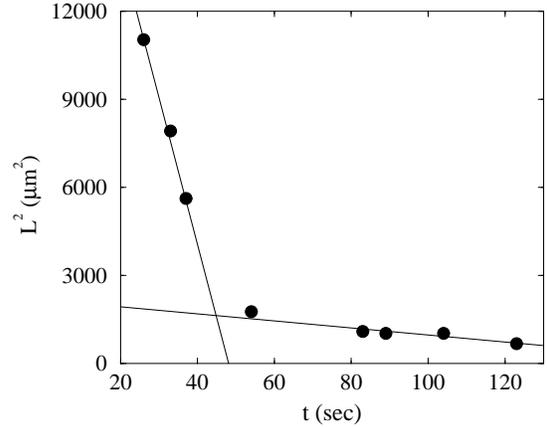}\hspace{1.0cm}}
	\vspace{0.25cm}
	\caption{Squared length, $L^2$, of the pearled portion of a tube 
	as a function of time, $t$, after the pipette is brought to the
	vicinity of the vesicle (circles). The straight lines are fits 
	of the early and late evolutions to the equation $L^2= 2D(\bar t - 
	t)$. At early times $D\approx 200\,\mu m^2/sec$, while at 
	late times $D\approx 5\,\mu m^2/sec$.}
\end{figure}

We have measured $\kappa\approx (20 \pm 5)\,k_{B}T$ using the
pipette aspiration technique \cite{evans:97}. Structures with radii 
smaller than optical resolution ($\sim 0.2\,\mu m$) were observed in our
experiments, giving
a lower bound of $H_0\gtrsim 10\,\mu m^{-1}$ on the spontaneous 
curvature induced by the polymer. Finally, $a$ can be estimated as 
the radius of gyration of a polymer performing a two dimensional 
random walk on the membrane. Since the hydrophilic backbone is in 
water, we assume a good solvent in semi-dilute conditions, giving
a radius between 40 and $80\,nm$.

We consider very long, cylindrically symmetric vesicles and ignore the 
existence of the end caps, since the length of most of the 
experimental tubes is larger than their radii by two orders of 
magnitude.  To find the equilibrium configuration of the system, the
free energy (\ref{eq:free}) was minimized with respect to the vesicle 
shape as well as the {\em local} polymer density. This was done under the 
constraints of constant vesicle volume, membrane area and total number 
of polymer molecules.


Ideally, the equilibrium configuration of the system would have a 
homogeneous polymer distribution and a curvature $H=\rho H_0$ 
everywhere.  Such a configuration minimizes the energy and maximizes 
the entropy simultaneously.  This is indeed possible for a range of 
values of $\rho H_0$. For example, a vesicle with volume-to-area ratio
$\lambda$ can have 
the shape of a cylinder of radius $2\lambda$ and curvature 
$H=1/(4\lambda)$. Another shape, having the same value 
of $\lambda$, is a chain of identical spheres of radius $3\lambda$ 
connected by infinitesimally narrow necks.  In this case
$H=1/(3\lambda)$ everywhere. In fact, the 
vesicle's curvature may take any intermediate value, 
$1/(4\lambda)<H<1/(3\lambda)$, because for each of these curvatures 
there corresponds a Delaunay shape \cite{Eells:87}, with the same 
value of $\lambda$. On the other hand, it is not possible to 
construct a shape of constant curvature for $H>1/(3\lambda)$ or for 
$H<1/(4\lambda)$.

In our experiment the polymer adsorbs onto the membrane gradually.  
Therefore, at the very early stages of the experiment 
$\rho<1/(4\lambda H_0)$, and the shape of constant 
curvature which minimizes the energy is a cylinder of radius 
$2\lambda$ and curvature $H=1/(4\lambda)$. This is consistent with our
experimental observations. In principle, we also have 
to consider possible inhomogeneities in membrane curvature. However, our 
calculations show that for small $\rho$ such inhomogeneities only
increase the free energy.

At intermediate stages of the experiment the polymer concentration is 
in the range $1/(4\lambda H_0)\leq\rho\leq 1/(3\lambda H_0)$, and long 
vesicles are expected to have a Delaunay shape.  Indeed, the shapes of 
vesicles we observe at intermediate stages of the experiment are 
similar to Delaunay shapes. Such pearled shapes have already 
been observed \cite{Ringsdorf:91}, and 
the importance of Delaunay shapes for this system has been discussed 
in the literature \cite{Deuling:77,Zhong:89}.

The situation becomes more interesting at late stages of the 
experiment when $\rho$ exceeds the value $1/(3\lambda H_0)$.  In this 
case, the best Delaunay shape is a chain of identical spheres of 
radius $3\lambda$ and curvature $H=1/(3\lambda)$.  However, a detailed 
calculation (to be presented elsewhere) shows that a configuration 
consisting of a chain of small spheres connected by infinitesimal necks 
to each other and to one large sphere has the lowest free energy.

The small spheres have polymer density $\rho_1$ and radius $r_1=1/(\rho_1 H_0)$.
The energy of this subsystem vanishes, 
since its curvature is $H_1=\rho_1 H_0$. The large sphere has radius 
$r_2$ and polymer density $\rho_2$, and plays the role of a reservoir
for the excess volume and polymer molecules. The ratios $r_2/r_1$ and 
$\rho_2/\rho_1$ can be calculated numerically as functions of the 
average polymer density, the area of the system and its volume.  For 
the values of the model parameters discussed above we obtained 
$r_2/r_1>10$ and $\rho_2/\rho_1<0.3$.  This ratio of the radii is 
consistent with the experimental configurations seen at long times, 
where chains of very small spheres coexist with few very large 
spheres, all connected by narrow necks (see Fig.\ 1d).

The strong inhomogeneity in polymer concentration predicted by the 
theory, is particularly interesting because it may show a qualitative 
difference between predictions of the local SC model and those of ADE 
models.  The latter does not differentiate between polymer molecules and 
lipids; i.e., exchanging polymers with a larger number of lipids of 
the same total area does not change the energy.  One can always use 
such exchanges to turn an inhomogeneous polymer distribution into a 
homogeneous one of higher entropy without changing the energy of the 
vesicle.  Hence, ADE models predict a homogeneous polymer 
distribution.  In the local SC model, inhomogeneities in polymer 
distribution induce the same undesirable decrease in the entropy of 
mixing. However, inhomogeneities lower the energy, and our 
calculations show that for reasonable parameter values, the reduction 
in energy does lead to sizable inhomogeneities.  
We intend to measure the polymer concentration on the membrane to 
test this interesting theoretical prediction.

We thank L. Jullien, for his help and encouragement, and
R. Lipowsky, E. Moses and S. 
Safran for useful exchanges. This research was supported by The Israel Science
Foundation - Recanati and IDB Group Foundation and The Minerva Foundation. V.F.\
acknowledges support from The Research Council of Norway
(NFR).

\vspace*{-3 mm}


\end{document}